\documentclass[10pt]{iopart}
\usepackage{graphicx}
\usepackage{latexsym}
\usepackage{dcolumn}
\usepackage{bm}
\usepackage{color}
\usepackage{amssymb}
\usepackage[normalem]{ulem}
\usepackage{mathtools}
\usepackage[utf8]{inputenc} 
\usepackage[T1]{fontenc}

\newcommand{\be}{\begin{equation}}
\newcommand{\ee}{\end{equation}}
\newcommand{\bea}{\begin{eqnarray}}
\newcommand{\eea}{\end{eqnarray}}
\newcommand{\bse}{\begin{subequations}}
\newcommand{\ese}{\end{subequations}}

\renewcommand{\e}{\epsilon}

\renewcommand{\comment}[1]{}

\begin{document}

\paper[]{Mean-field theory of the DNLS equation at positive and negative absolute temperatures}
\date{\today}

\author{Michele Giusfredi$^{1,2,3}$, Stefano Iubini$^{2,3}$, Antonio Politi$^{2,4}$, Paolo Politi$^{2,3}$}

\address{$^{1}$ Dipartimento di Fisica e Astronomia, Universit\`a di Firenze,
via G. Sansone 1 I-50019, Sesto Fiorentino, Italy}
\address{$^{2}$ Istituto dei Sistemi Complessi, Consiglio Nazionale
delle Ricerche, via Madonna del Piano 10, I-50019 Sesto Fiorentino, Italy}
\address{$^{3}$ Istituto Nazionale di Fisica Nucleare, Sezione di Firenze, via G. Sansone 1 I-50019, Sesto Fiorentino, Italy}
\address{$^{4}$ Institute for Complex Systems and Mathematical Biology
           University of Aberdeen, Aberdeen AB24 3UE, United Kingdom}
\ead{michele.giusfredi@unifi.it,stefano.iubini@cnr.it,\\a.politi@abdn.ac.uk,paolo.politi@cnr.it}

\begin{abstract}
The Discrete Non Linear Schr\"odinger (DNLS) model, due to the existence of two conserved quantities,
displays an equilibrium transition between a homogeneous phase at positive absolute temperature and a localized
phase at negative absolute temperature.
Here, we provide a mean-field theory of DNLS 
through a suitable approximation of the grandcanonical partition function
which makes it factorizable and can be used to describe the equilibrium state 
at positive temperatures as well as the metastable state at negative temperatures.
By comparing our mean-field results with numerically exact ones, we
show that this approximation is good-to-excellent in the whole grandcanonical phase diagram.
Explicit approximate expressions for equilibrium observables are provided in the high-temperature limit.
Our theory represents a clear advancement over the model that neglects the interaction between sites.
\end{abstract}

\noindent{\bf Keywords:} Negative temperature; Mean-field theory; Phase transitions

\submitto{Journal of Statistical Mechanics: theory and experiment}

\maketitle

	\section{Introduction}

The Discrete Non Linear Schr\"odinger (DNLS) equation is around fifty years old, but
it continues to attract the interest of researchers~\cite{kevrekidis09,Flach2018_PRL,arezzo22,giachello25,Kabat26} 
for at least two reasons.
On the one hand, it has applications in a variety of different fields,
ranging from the foundations of statistical mechanics~\cite{Baldovin_2021_review} 
to solid-state physics~\cite{tsironis2025DNLS}.
On the other hand, its nonlinear and non integrable character 
is the basis of interesting non-trivial properties:
from a dynamical point of view, the existence of localized breathers~\cite{flach98},
and from a statistical point of view, the presence of a negative-temperature ($T$) regime~\cite{Iubini2013_NJP}
in a large part of its phase diagram.
The two properties are strictly related, because the negative$-T$ phase
is characterized by the spontaneous emergence of breathers, which are
in principle metastable~\cite{rumpf09}
(the ``equilibrium'' state corresponds to a single breather sitting
on top of an infinite-temperature background~\cite{ebrahimi25}).
However, the rising of breathers is an extremely slow process: for large and negative 
temperatures a high barrier hinders their divergence~\cite{IP25}; moreover the dynamics itself is
extremely slow~\cite{PRL_DNLS}, due to the
spontaneous appearance of an adiabatic invariant~\cite{PPI2022_JSTAT}.
As a result, the asymptotic state may be effectively unachievable~\cite{ebrahimi25},
making the study of the homogeneous, metastable state of special interest.


According to Machlup's criterion, the existence of a negative$-T$ phase is a general property of statistical systems
displaying a finite energy density at infinite temperature~\cite{machlup75}. 
In the DNLS case, this occurs because of the existence of two conserved quantities bounded from below: the energy and the mass (or norm).
The thermodynamics of DNLS  was studied by Rasmussen et al.~\cite{Rasmussen2000_PRL}
in the grandcanonical ensemble through the transfer integral operator. This approach allowed to identify an infinite-temperature line of finite
energy densities separating homogeneous states at positive temperature from localized negative$-T$ states.
In the positive-$T$ region, the formal expression of the grandcanonical partition function is usually of unwieldy usage for practical 
purposes, while for negative temperatures it is formally ill-defined due to the divergence of the integral over the grandcanonical distribution.
This scenario motivated further studies of simplified models sharing some basic features with the full DNLS equation.

A dozen years ago it was introduced a purely stochastic 
model~\cite{Iubini2013_NJP,JSP_DNLS,GIP21,Szavits2014_PRL} recently renamed C2C model, because it
displays a condensed phase induced by two conservation laws. In fact, it shares with the DNLS the
existence of two conserved quantities bounded from below: the energy and the mass (or norm).
The absence of interaction energy between neighbouring sites in the C2C model  makes it possible
to derive some rigorous results~\cite{GIP24,GIPP25}.
Furthermore, it helped establishing 
a connection with a wider class of stochastic models~\cite{Gradenigo2021_JSTAT,Gradenigo2021_EPJE}, 
the so called Zero Range Processes~\cite{evans05r}.
The C2C model gives an accurate thermodynamical description of the DNLS  in proximity of the transition line
between homogeneous and localized/condensed phase ($T=\infty$), but it fails when one moves away
from this line because  C2C disregards the coupling between different sites.

More recently, the DNLS model was studied through some type of mean-field (MF) approach in the
large connectivity limit where each site interacts with all the others~\cite{arezzo22}.
Another approach has recently employed transfer integral techniques to develop an equilibrium description in terms
of interacting Rayleigh-Jeans modes beyond the weakly nonlinear limit~\cite{Kabat26}.

%
%

In this paper we introduce a MF description of the DNLS, where the MF approximation is applied
exclusively to the mass variables, while
treating the phases exactly and keeping nearest-neighbors interactions. 
The approximation is based on a tweak of the interaction term: we replace the product of variables $x_n x_{n+1}$
on neighbouring sites,
with the product between the variable on a given site with its statistical average 
along the system, $x_n \langle x_n\rangle$.
As a result, the grandcanonical partition function can be factorized, thereby obtaining 
explicit formulas and reaching a better comprehension of some properties of the DNLS model. 

As we are going to see, this allows to obtain a consistent
and semiquantitatively correct description in the whole phase diagram,
including the $T=0$ curve and the (metastable) negative $T$ region. 
In fact, as shown in Refs.~\cite{Gradenigo2021_JSTAT,Gradenigo2021_EPJE},
the equilibrium state of DNLS at negative $T$ can not be described in the grandcanonical ensemble,
but at small, negative $T$ it is possible to give a consistent description of the metastable state,
which is stable on very long time scales.
Therefore, our MF approximation is not only asymptotically exact in proximity of the transition line,
but also allows obtaining explicit analytical expressions which are applicable on both sides of the critical line.

In Sec.~\ref{sec.MMF} we present the model and the method. 
The main results of the MF approximation are discussed, distinguishing
between positive (Sec.~\ref{sec:pos}) and negative (Sec.~\ref{sec:neg}) 
temperatures, because in the latter case it 
is necessary to introduce a cutoff in the mass integral which defines the partition function.
All the details of the calculation and of the simulations 
are deferred to the appendices.

\section{The Model and its Mean Field Theory}
\label{sec.MMF}

We start by defining the DNLS model and recalling the most important properties of its phase diagram.
In one dimension, the Hamiltonian is
\be
H = \sum_n \left[ |z_n|^4 + 2J(z_n z_{n+1}^* + z_n^*z_{n+1}) \right ]
\label{eq.H0}
\ee
where $z_n$ is a complex variable describing the state of the system in the site $n$ and the
parameter $J\geq 0$ modulates the hopping energy.\footnote{Negative values of $J$ can be mapped to 
positive ones via the gauge transformation $\phi_n\to \phi_n+\pi n$~\cite{kevrekidis09}.} 
The corresponding Hamilton equations are
\be
i\dot z_n = - 2 |z_n|^2z_n - J(z_{n-1} + z_{n+1}) \, .
\ee
In this paper, it is convenient to express the equations in terms of action-angle variables, 
defining $z_n = \sqrt{c_n}{\rm e}^{i\phi_n}$, where $c_n\ge 0$ can be interpreted as a local mass. 
The Hamiltonian becomes
\be
H = \sum_n\left[ c^2_n + 2J \sqrt{c_n c_{n+1}} \cos(\phi_n - \phi_{n+1})\right] \equiv 
N(h_{nl} + h_{int}) = Nh,
\label{eq.H}
\ee
where $h_{nl}$, $h_{int}$, and $h$ are the nonlinear, the interaction, and the total energy densities,
respectively. Consistently, the dynamical equations are
 $\dot c_n = -\partial H/\partial \phi_n$ and
$\dot \phi_n = \partial H/\partial c_n$.
The mass
\be
A = \sum_n c_n \equiv Na ,
\label{eq.A}
\ee
as well as the energy, is conserved.

The ground state is characterized by uniform masses ($c_n =a$) and alternating phases
($\phi_{n+1} =\phi_n + \pi$). The ground state energy is therefore
\be
h_{GS} = a^2 -2Ja .
\ee
The transition line between positive and negative temperatures is characterized by
an exponential distribution of masses and a random distribution of phases, which implies
\be
h_c = 2a^2.
\ee
The negative$-T$ region appears above the critical line and its properties are 
not yet fully understood~\cite{GIP21,Gradenigo2021_JSTAT,IP25}:
they depend on the size of the system, on the statistical
ensemble describing it, and on the time scales over which we observe the system,
because the homogeneous phase is actually metastable above the critical line.

In the limit case $J=0$, the interaction energy $h_{int}$ vanishes and the angles do not
obviously contribute. Actually, this corresponds to the so-called C2C model; given the absence
of interactions, an evolution can be implemented via a local
stochastic dynamical rule that ensures conservation of both mass and energy~\cite{JSP_DNLS}. 

Within the grand canonical formalism, the equilibrium properties of the DNLS model can be extracted from the partition
function
\be
Z(\beta,\mu) = \int_0^\infty \prod_n dc_n \int_0^{2\pi} \prod_n d\phi_n
{\rm e}^{-\beta(H+\mu A)},
\ee
where $\beta$ is the inverse temperature, $\mu$ is the chemical potential, while
the energy $H$ and the mass $A$ are given by Eqs.~(\ref{eq.H}-\ref{eq.A}).
The function was evaluated in Ref.~\cite{Rasmussen2000_PRL},
determining the leading eigenfunction of a suitable transfer integral operator.
However, the final expressions are not very handy for the extraction of useful information.
Altogether, the main obstacle is that for $J \ne 0$, the interaction energy prevents a factorization of the integrals.

Here, we show that a suitable mean field approximation restores the factorization.
Our method consists in rewriting the product of 
masses appearing in the neighbouring-site coupling term, see Eq.~(\ref{eq.H}), as
\be
\sqrt{c_n c_{n+1}} \simeq q \sqrt{c_n},
\ee
where $q\equiv \langle \sqrt{c_n}\rangle$ is a statistical average
to be determined autoconsistently with $Z$. Within this approximation,
the Hamiltonian $H$ is rewritten as
\be
H_{MF} = \sum_n \left[ c^2_n +  2qJ\sqrt{c_n} \cos(\phi_n - \phi_{n+1}) \right]
\ee
and the partition function becomes the product of single-site terms,
$Z_{MF} = z^N$, with
\be
z(\beta,\mu) = \int_0^\infty dc \int_0^{2\pi} d\varphi 
{\rm e}^{-\beta(c^2 + 2qJ\sqrt{c} \cos\varphi) + mc},
\label{eq:z}
\ee
where $\varphi$ is the phase difference between neighbouring sites, and
$m\equiv \beta\mu$.

The integral over the angle $\varphi$ can be easily calculated, introducing
the zero-order modified Bessel function,
\be
    z(\beta,\mu) = 2\pi \int_0^\infty d c\, \exp \left( - \beta c^2  + m c\right) I_0( 2 \beta qJ \sqrt{c} ), 
    \label{eq_z1}
\ee
where $q$, as anticipated, must be determined autoconsistently through the relation
\be
q \equiv \langle \sqrt{c}\rangle = \frac{2\pi}{z} \int_0^\infty d c\, \sqrt{c} \exp\left( - \beta c^2  + m c\right) I_0( 2 \beta qJ \sqrt{c} ). 
\ee
Similar expressions can be also derived for the mass density $a=\langle c\rangle$ and the
energy densities $h_{nl}= \langle c^2 \rangle$ and $h_{int} = \langle 2qJ\sqrt{c}\cos\varphi\rangle$. 
Without loss of generality, from now on we will assume $J=1$, because $J$ can be scaled out so long as it is strictly larger than 0. 

If $\beta > 0$ the convergence of the integrals appearing in the definition of the various observable is ensured, while for $\beta < 0$
the integrand diverges at large $c$, signaling the well known instability due to the
appearance of large peaks (breathers). For this reason, negative temperatures require a special treatment and we have thereby
devoted an entire section to their discussion. In a few words, while  negative-temperature localized states in the DNLS model are formally well-defined
only in the microcanonical ensemble~\cite{Gradenigo2021_JSTAT,Gradenigo2021_EPJE,giachello25}, homogeneous, metastable states
have been recently shown to admit a consistent description in terms of a regularized grand canonical theory~\cite{IP25}.
In this paper we will restrict ourselves to this class of negative-temperature states.

Independently of the sign of $\beta$, it is convenient to introduce the smallness parameter 
$w=1/(\beta\mu^2) =\beta/m^2$ ($m=\beta\mu\to - 1/a$ for $\beta\to 0$), which allows deriving simple and accurate
formulas in the small $|\beta|$ region (see~\ref{app.exp} for the derivation of the expressions for the most relevant observables).

For instance, at leading order in $w$, the following expressions hold 
across the critical line, i.e., for both positive an negative temperatures:
\be
q = \frac{ \sqrt{\pi} }{ 2 \sqrt{|m|} } - \frac{7 \sqrt{\pi } w}{8 \sqrt{|m|}} + O\left( w^2\right) 
\label{eq.qM}
\ee

\be
a = \frac{1}{|m|}-\frac{4 w}{|m|}+ O\left( w^2\right) 
\label{eq.awM}
\ee

\be
    h_{nl} = \frac{2}{m^2}-\frac{20 w}{m^2} + O\left( w^2\right) 
\label{eq.hnlwM}
\ee

\be
    h_{int} =   -\frac{\pi }{2} w+ O\left( w^2\right) .
\label{eq.hnlwM2}
\ee

\section{Positive temperatures}
\label{sec:pos}

\begin{figure}
        \centering
        \includegraphics[width=1\columnwidth]{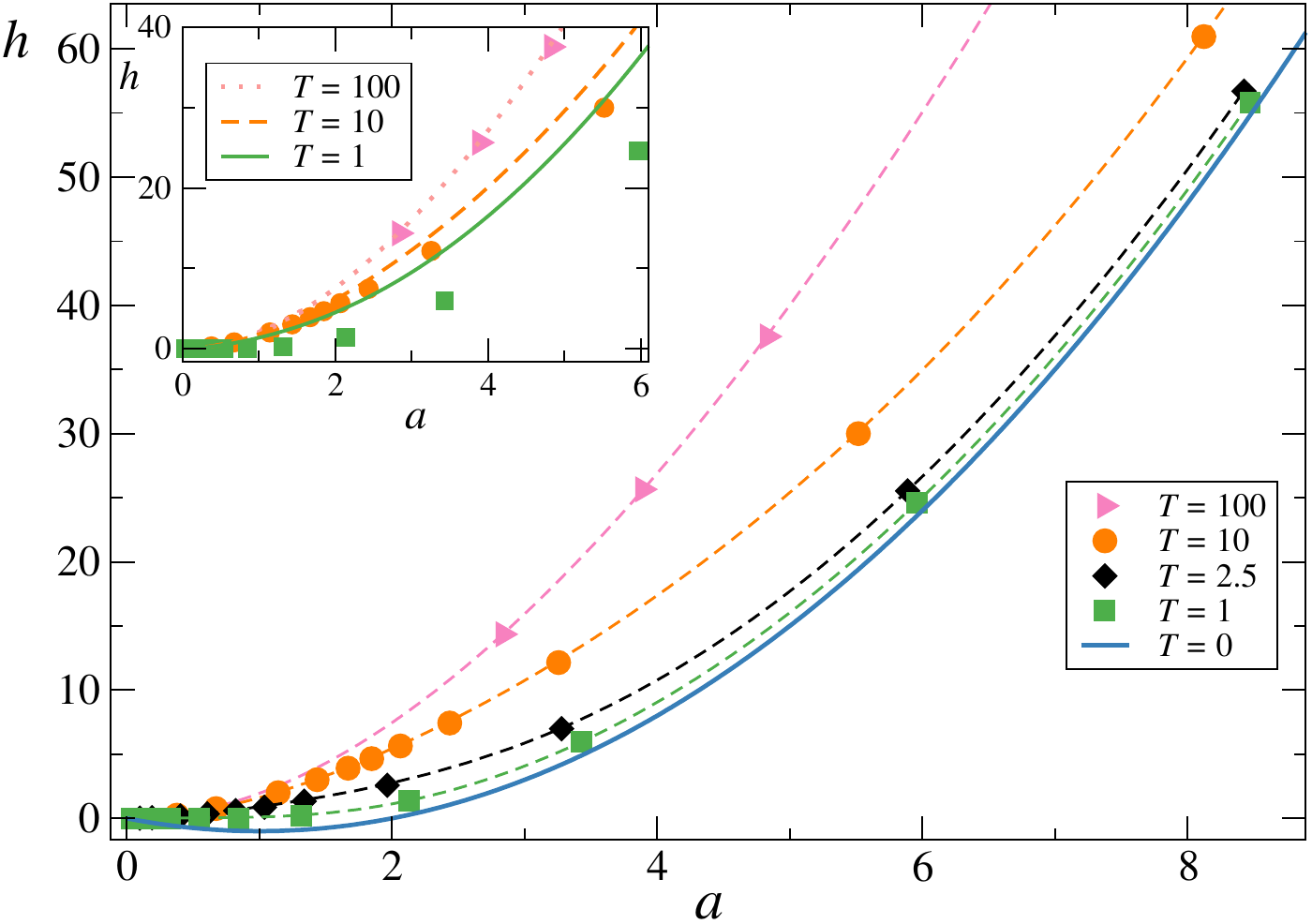} 
        \caption{
		Main: Comparison between MF theory (curves) and exact results obtained via
		grandcanonical simulations (symbols). 
		Each curve corresponds to a different $T=1/\beta$, decreasing from top to bottom
		according to the legenda. At $T=0$, the ground state curves for DNLS and its MF
		approximation are known exactly and they coincide (full line).
		Inset: Comparison between C2C model (curves) and exact results (symbols).
		The agreement is acceptable only for very large $T$.
		The size of the system is $N=100$ and details about simulations are given
		in~\ref{app.tech}.
	}
        \label{fig:ahpb}
    \end{figure}

In the positive$-T$ regime, the equilibrium state can be equivalently described by the microcanonical
and the grandcanonical ensemble~\cite{Gradenigo2021_JSTAT,Gradenigo2021_EPJE}. 
Within the MF approximation, this means
that the single-site partition function, see Eq.~(\ref{eq:z}), is well defined and finite
for any $\beta \ge 0$ and for any $m$ (i.e., for any $\mu$).

In Fig.~\ref{fig:ahpb} we compare virtually exact results obtained via grandcanonical simulations (symbols)
with MF curves (main panel) and with the C2C model (inset),
see \ref{app.tech} for numerical details.
These data show that the isothermal curves $h(a)$ are exceptionally well reproduced by the MF theory, down to $T=0$:
in fact, the ground state energy has the same analytical expression, $h_{GS}(a) = h^2 -2Ja$.
The agreement is not as good for the C2C model, as shown in the inset of the same figure:
already for $T=10$, the isothermal curve of C2C compares very poorly with exact results.

\begin{figure}
        \centering
        \includegraphics[width=1\columnwidth]{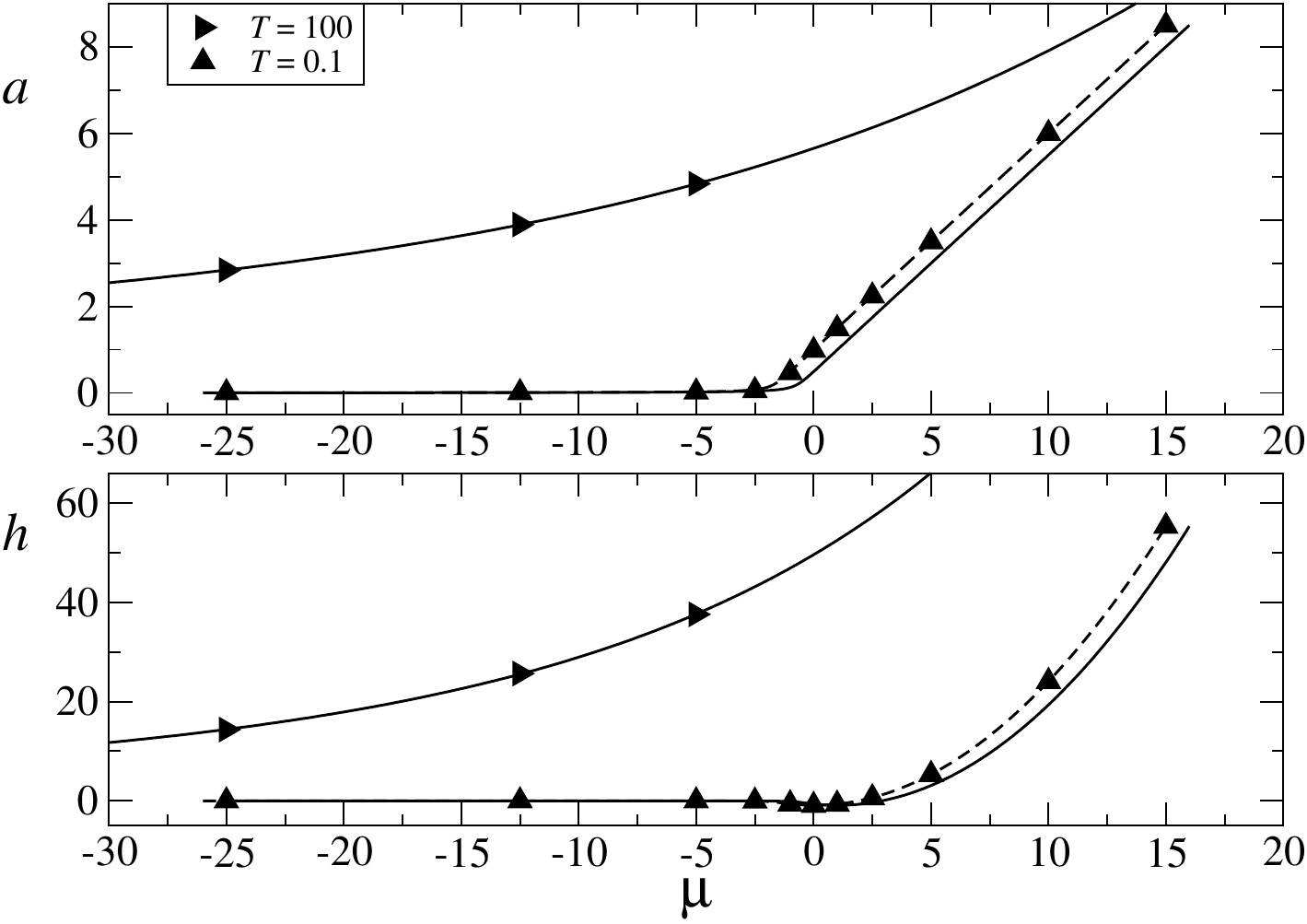} 
	    \caption{Same as in Fig.~\ref{fig:ahpb} main, but we now compare MF and exact numerical results
	    in the planes $(\mu,a)$ and $(\mu,h)$ rather than in $(a,h)$.
	    In this way a horizontal shift at low temperature is evident, due to a wrong determination 
	    of the chemical potential in the MF theory. Symbols: exact results from grandcanonical
	    simulations. Full lines: MF approximation. Dashed lines: MF approximation corrected
	    for the chemical potential: $\mu\to\mu -1$.
        }
	    \label{fig:muahpb}
    \end{figure}

In Fig.~\ref{fig:muahpb}, the validity of the MF theory is also tested
in the planes $(a,\mu)$ and $(h,\mu)$.
There, we see that the agreement with the numerical data is very good at high $T$, while 
it worsens with decreasing $T$. 
Remarkably, it appears that the deviation essentially amounts to a shift of the chemical potential
by one unit, as it can be appreciated by looking at the dashed curve, which corresponds to $\mu-1$.
As shown in~\ref{app.T0}, the shift is a rigorous property at zero-temperature, and it is essentially
irrelevant while approaching the transition line for finite $a$, since $\mu$ diverges therein,
so that the curves for $\mu$ and $\mu -1$ practically coincide.

We now discuss the relative role of $h_{nl}$ and $h_{int}$ for a double reason:
because it allows a further test of the MF approximation and because it is
an aspect of DNLS that has been little considered in its studies.
From the literature, we know only that $h_{int}$ vanishes on the critical line, 
which is the reason why C2C model is statistically equivalent to DNLS in its proximity.
Here below we try to understand more quantitatively how the energy distributes
between the two terms, both in the exact DNLS theory and in its MF approximation.

\begin{figure}
        \centering
		\includegraphics[width=1\columnwidth]{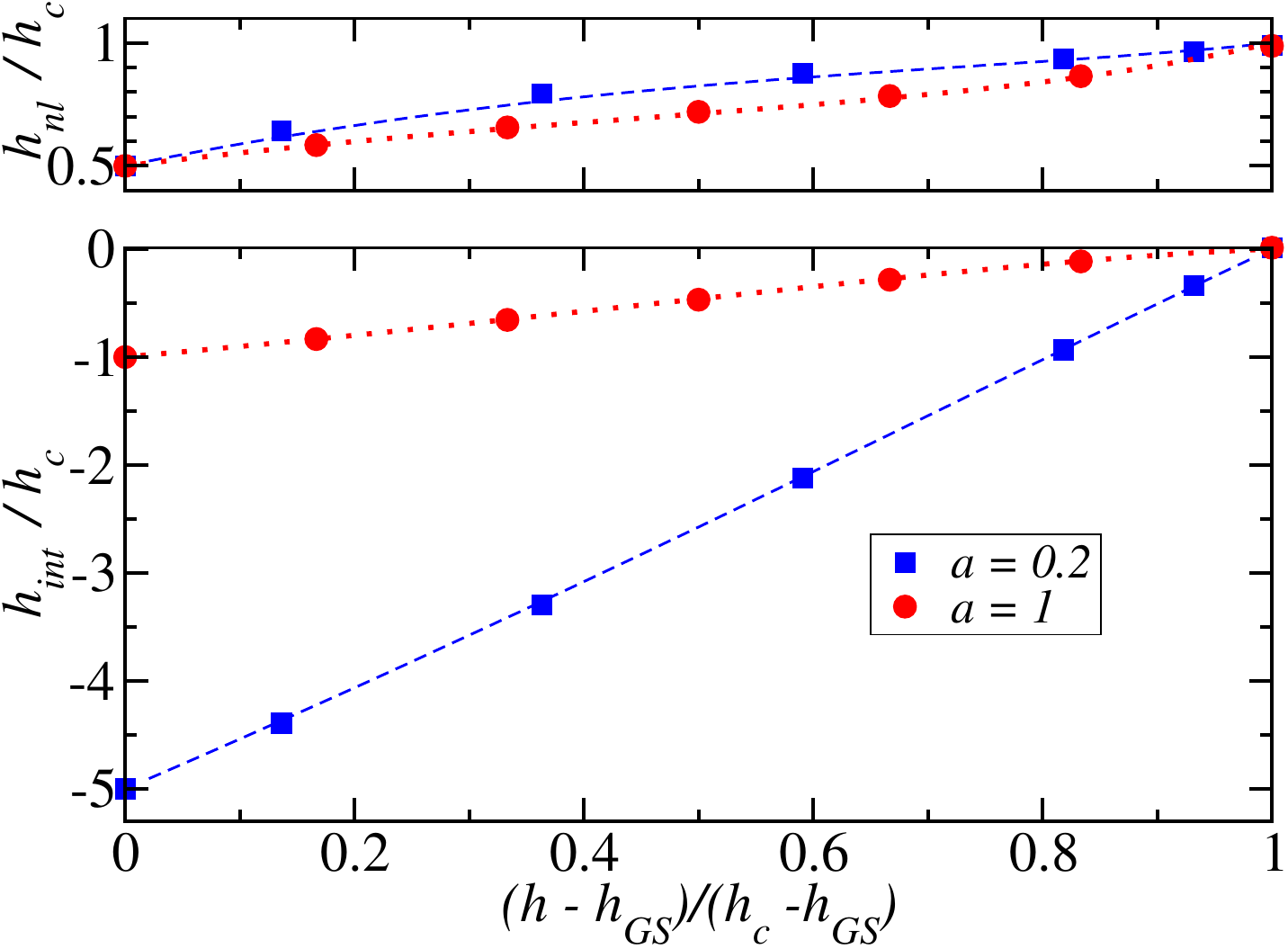}
		\caption{The relative contributions $h_{int}$ (lower panel)
and $h_{nl}$  (upper panel)
        are presented versus
		the total energy $h$ suitably rescaled (see the main text).
	Symbols are the  exact results  from microcanonical simulations, as described in Appendix A. 
		Dashed ($a=0.2$) and dotted ($a=1$) lines are the MF results.
}
		\label{fig:hlin}
        \end{figure}

In Fig.~\ref{fig:hlin}
we present the dependence of $h_{nl}$ and $h_{int}$ \textit{vs} the total energy $h$, for two values of
the mass density.
In order to facilitate the comparison, the total energy is suitably shifted and scaled
in such a way that
the zero of the horizontal axis corresponds to $T=0$, while 1 corresponds
to infinite temperature. This corresponds to plot 
$x=(h-h_{GS})/(h_c -h_{GS})$.
Similarly, the two energy terms are scaled by the critical energy;
so we plot $y_{nl,int}=h_{nl,int}/h_c$.
Using the new variables, we have $y_{nl}(0)=\frac{1}{2}$ and $y_{nl}(1)=1$,
while $y_{int}(0)=-1/a$ and $y_{int}(1)=0$.
	We stress that MF results (lines) compare very well with exact results (symbols).
It is also remarkable that both functions $y_{nl,int}(x)$ are close to linear.

It is not possible to derive analytically the behavior of $y_{nl,int}(x)$ within the MF theory,
because such theory works with the grandcanonical variables $\beta,\mu$ and their relations
with $a$ and $h$ can be obtained numerically, but no closed-form solutions are available.
However, we can use the expansions in $w$, which is what we are going to do now.
For this purpose we have identified a quantity that for small $w$ depends on a single parameter,
namely $m$ (see~\ref{app.Hlnl}),
\be
\label{eq:R}
R= \frac{h_c - h_{nl}}{h_c -h} \; .
\ee
This is the ratio between the distance of the nonlinear energy
from the $\beta=0$ value (for the same mass density) and the distance of the full energy, again from
the $\beta=0$ value.

In the positive-temperature region both the numerator and the denominator defining $R$ are positive. Moreover, since $h_{int} \le 0$, 
the numerator is smaller than the denominator so that $0 \le R \le 1$.
However, it is not straightforward to identify the locations in parameter space where the limit values
can be achieved.
In fact, for $R$ being equal to 1, it is necessary that $h_{int}=0$, which occurs when
the denominator and the numerators are both equal to 0. Furthermore, $R=0$ requires 
$h_{nl}=2a^2$, which again occurs when the denominator vanishes.

        \begin{figure}
        \centering
		\includegraphics[width=1\columnwidth]{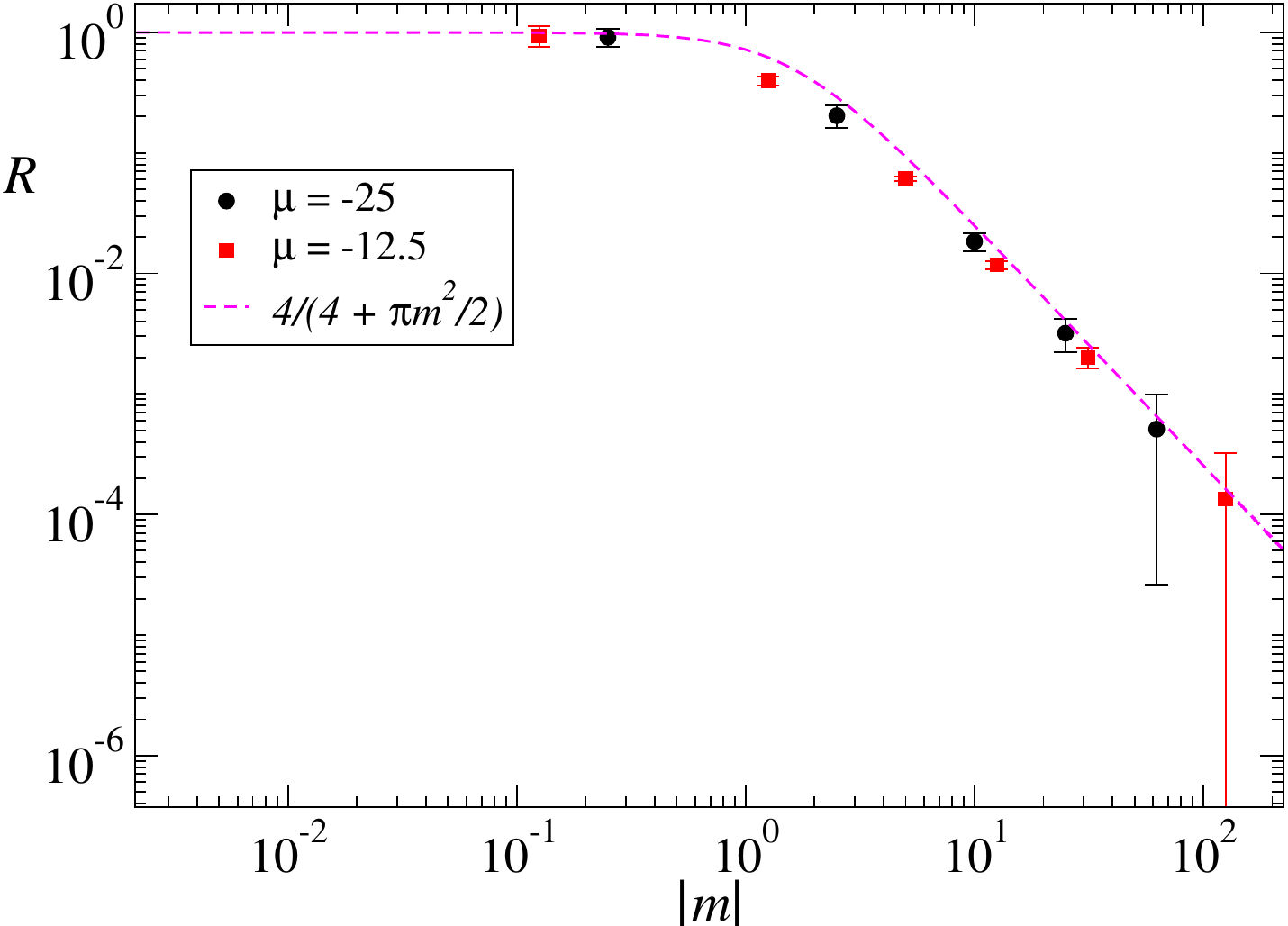}
		\caption{
		Dependence of the parameter $R= (h_c-h_{nl})/(h_c-h)$  versus $|m|$. Symbols refer to numerical simulation of the full
		 DNLS model, while the dashed
		line refers to the analytical prediction of the MF theory.
		}
		\label{fig:Hlnl}
    \end{figure}

MF analysis helps clarifying the issue, as shown in Fig.~\ref{fig:Hlnl}, 
where we plot $R$ for different values of $\mu$ and $\beta$, comparing numerical results
for the full DNLS model (symbols) with the analytical curve $R=4/(4 + \pi m^2/2)$, found within the
MF approximation, in a $w-$expansion. At the lowest order in $w$, $R$ is only function of $m$, not of $w$.
It is worth remembering that $m=\beta\mu$ is equal to $-1/a$ on the critical curve $\beta=0$. 
On the other hand, if we fix $\mu$ and vary $\beta$, the limit of small $\beta$ corresponds to small $m$. 
Let us start from
this limit, because for $\beta\to 0$, $h_{nl}\to 0$ and $R\to 1$: this is what we observe in Fig.~\ref{fig:Hlnl}.
We can also monitor the regime $|m|\ll 1$, provided that $w$ is small. On the critical curve, this simply
implies $a\ll 1$ and in this limit $R\simeq a^2$, which means that for small mass density $a$, when $h$
approaches $h_c$, $h_{nl}$ is practically on the critical curve.

It is worth stressing that the approximation $R=4/(4 + \pi m^2/2)$ is not valid for vanishing $T$.
In this limit, $m$ diverges, which would imply a vanishing $R$, while in reality $R$ goes to a constant,
\be
\lim_{T\to 0}  R = \frac{a^2}{a^2 + 2a} .
\ee

\section{Negative-temperature homogeneous states}
\label{sec:neg}

States with energy density above the critical line $h_c(a)$ display negative temperatures and are expected to develop a thermodynamic
instability, resulting in the  condensation of a macroscopic peak~\cite{Rasmussen2000_PRL}. The theoretical description of this mechanism
requires a microcanonical formalism, while the grand canonical measure is formally ill-defined~\cite{Gradenigo2021_JSTAT,Gradenigo2021_EPJE}.
Recently it was shown that homogeneous states slightly above the critical line (i.e. with $|\beta|\ll 1)$ are metastable, in the sense that the 
characteristic time to develop the localization instability grows exponentially with $|T|$~\cite{IP25}. 
In this regime, an effective grand-canonical description can be developed 
in terms of an asymptotically unstable thermodynamic potential with a metastable region for low-amplitude states.
The fundamental properties of the effective potential were derived in~\cite{IP25} under the approximation that the hopping term of the DNLS
Hamiltonian is negligible with respect to the nonlinear one, thus removing any explicit phase dependence.
In the following, we generalize this derivation including phases, within the MF approximation.

\begin{figure}
        \centering
		\includegraphics[width=0.8\textwidth]{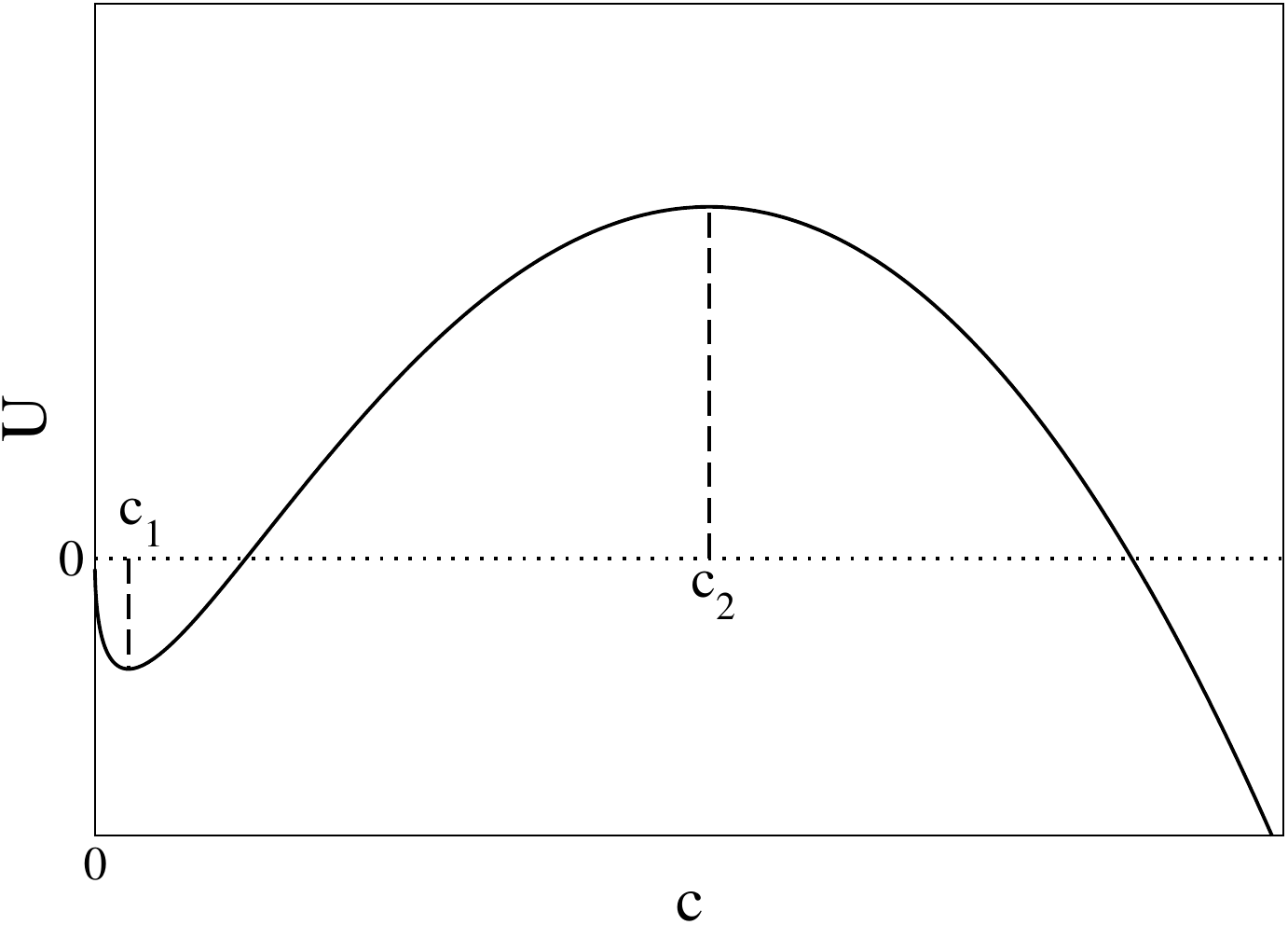}
		\caption{Sketch of the potential $U(c)$, Eq.~(\ref{eq:Uc}),
for $\phi=0$ and small, negative $\beta$.}
	\label{fig.potenziale}
\end{figure}

For $\beta <0$, we introduce a cutoff $c^*$ whose value is univocally determined by the
potential, see below. Said this, Eq.~(\ref{eq:z}) is modified as follows,
\be
z(\beta,m) = \int_0^{c^*} dc\, \int_0^{2\pi} d\phi\, \exp\left[-U(c,\phi;\beta,m)\right] \,,
\label{eq:zNT}
\ee
where we have rewritten the exponent as a two-dimensional potential having the form
\be
U(c,\phi;\beta,m) = \beta c^2 +  2\beta q \sqrt{c} \cos\phi - m c .
\label{eq:Uc}
\ee
Notice that the parameters $\beta$ and $m$ are both negative, therefore
giving a potential (\ref{eq:Uc}) whose shape is qualitatively plotted 
in Fig.~\ref{fig.potenziale} for $\phi=0$.

The theory of the metastable state proceeds as follows. We start determining the lowest
barrier, corresponding to $\phi=0$. This barrier (which is a function of $q$) determines $z$
and any thermodynamic average, including $q=\langle\sqrt{c}\rangle$ itself.
Once $q$ has been found self-consistently, we have a theory allowing to pass from the
grandcanonical variables $\beta,\mu$ to the microcanonical ones, $a,h$.
In the C2C model~\cite{IP25}, the procedure is similar but easier because
$c^*=m/2\beta$ is independent of $q$.

\begin{figure}
        \centering
        \includegraphics[width=1\columnwidth]{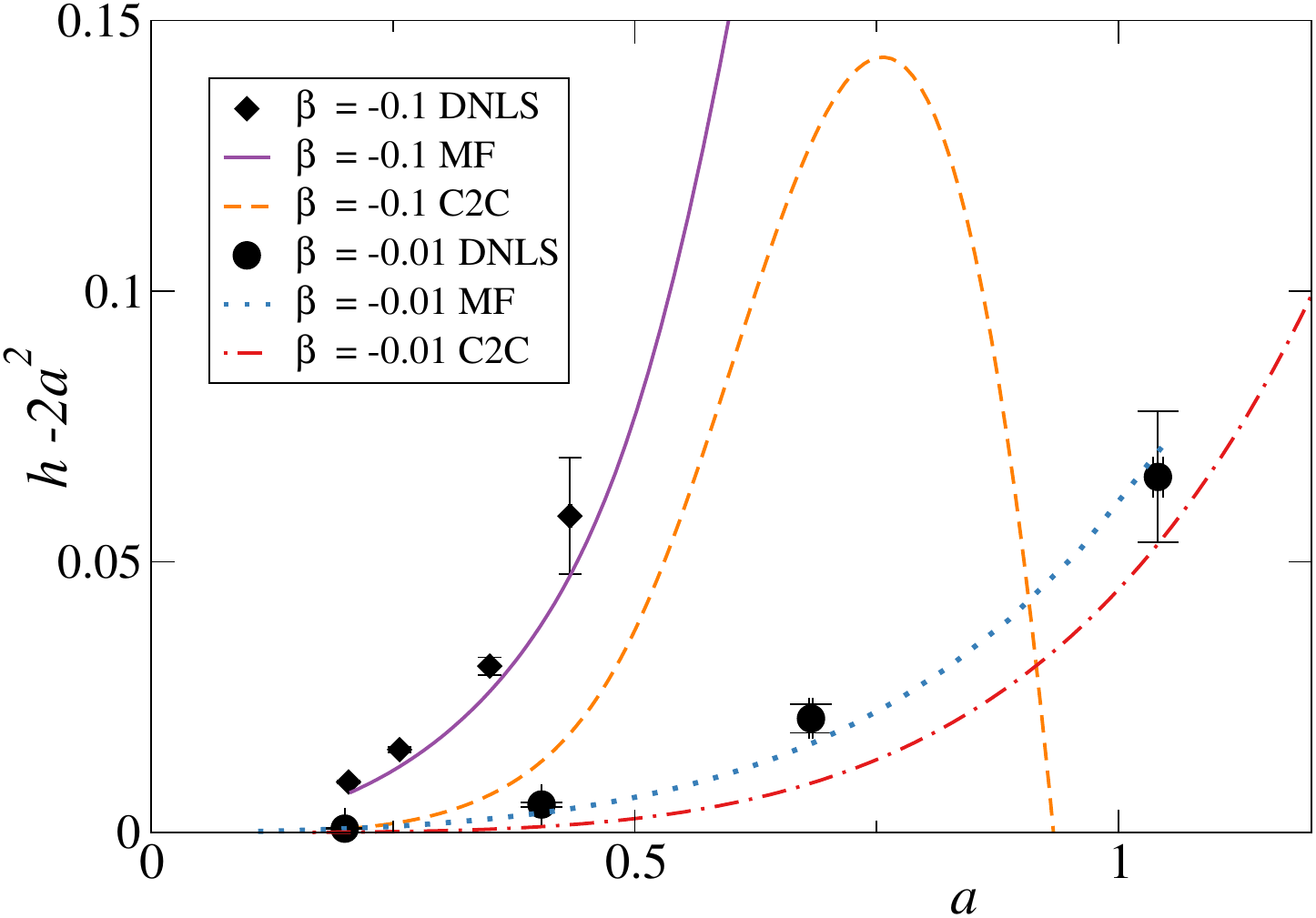}
            \caption{Comparison of exact numerical results (symbols)
            with MF predictions (full and dotted lines) and the C2C model
            (dashed and dotted-dashed lines).
            We make such comparison for two negative temperatures,
            $\beta=-0.1$ and $\beta=-0.01$:
            It is manifest we are in the negative$-T$ region, because $(h-2a^2)>0$.
            We also make evident the statistical error bar for the symbols, obtained
            as spatial and temporal averages of grandcanonical simulations.
            }
        \label{fig:mf6}
    \end{figure}

In Fig.~\ref{fig:mf6} we compare virtually exact numerical results for the DNLS model with
its MF approximation and with results for the C2C model, see the legend for more details.
Simulations of the DNLS equations have been performed employing negative-temperature 
Monte Carlo (grand-canonical) reservoirs, see~\ref{app.tech}.
We perform this analysis in the plane $(h-2a^2,a)$ to better highlight the small differences between the
energy and the critical energy. While the MF theory gives a relatively accurate approximation, the 
C2C model does not, even for the smallest (negative) inverse temperature, $\beta= -0.01$.
For $\beta=-0.1$, not only C2C compares badly with DNLS, it also displays a re-entrant behavior.

This is because for larger $a$ the confining character of the potential
is restricted to a progressively smaller interval of negative $\beta$ 
and the stability of the homogeneous state is quickly lost.
Outside this condition it is therefore no longer possible to give a grandcanonical description.
An inspection of potential (\ref{eq:Uc}) allows to be more precise,
distinguishing among small, intermediate, and large $c$ values:
$U(c) \approx -\sqrt{c},c,-c^2$, respectively.

The crossover from the first to the second regime occurs at $c_1 \simeq \beta^2 q^2 / m^2 $, while 
the crossover from the second to the third regime occurs at $c_2\simeq |m/\beta|$.
For the potential to be partially confining, it is therefore necessary that $c_2$ is
sufficiently larger than  $c_1$.
This implies $|m| \gg |\beta| q^{2/3}$, which, in the limit of small $\beta$, leads to 
$a\ll 1/|\beta|^{3/4}$. This is a necessary condition to have some confinement,
the condition to have stability on a meaningful timescale is much stronger, but it is
not possible to make predictions without studying the dynamics.

\begin{figure}
        \centering
             \includegraphics[width=1\columnwidth]{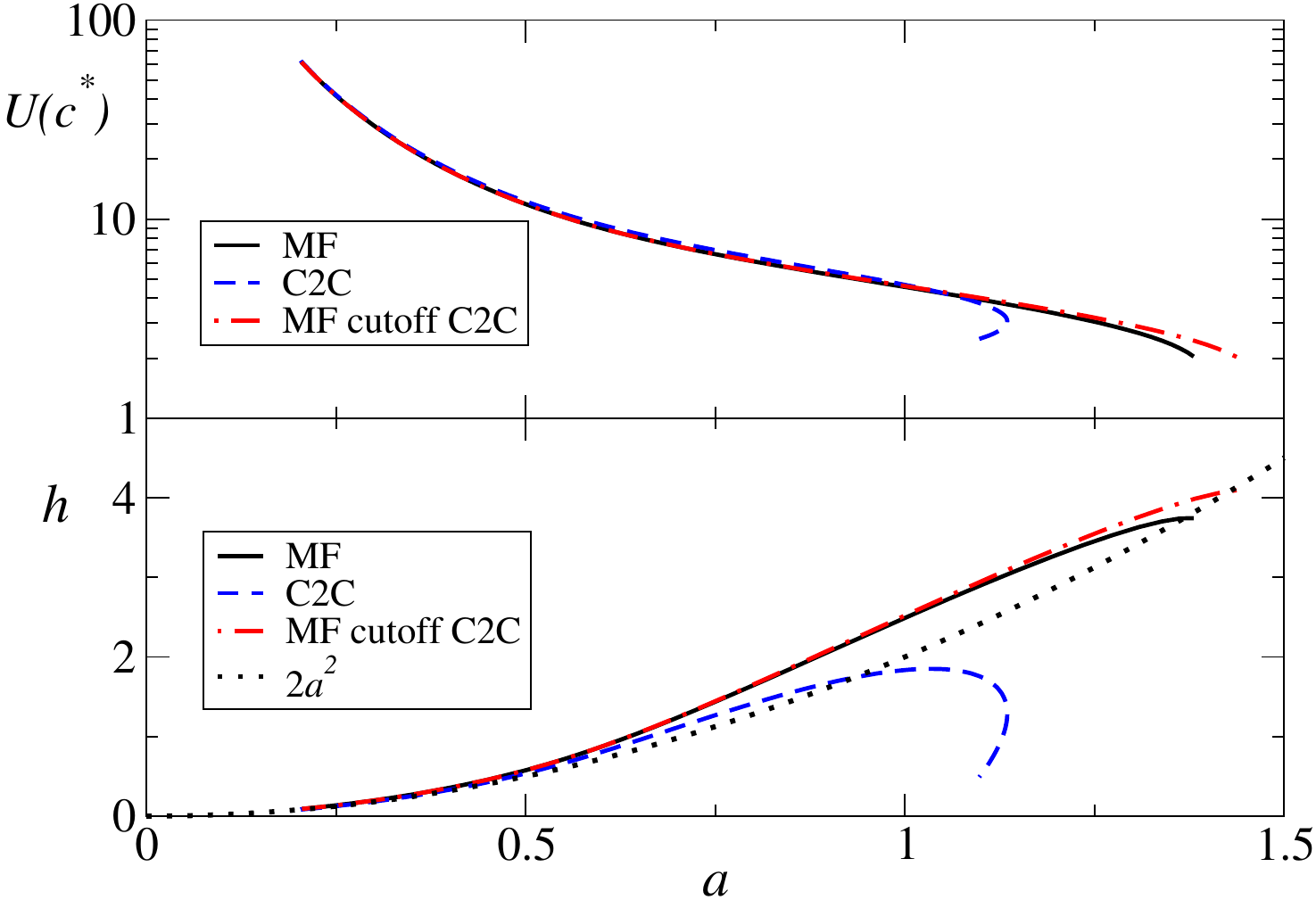}
            \caption{
                    Upper panel: Comparison of the height of the barrier for MF theory,
             C2C model, and the MF theory using the C2C cutoff.
             Lower panel: Same type of comparison, for the parametric curves $h(a)$.
	     Data refer to $\beta=-0.1$ and variable $\mu$ in the interval $(10,50)$.
            }
        \label{fig:mfnew}
    \end{figure}

The fact that MF theory allows for a better description of DNLS than C2C model is not because of a better
evaluation of the cutoff $c^*$ or of the energy barrier $U(c^*)$:
it is because the interaction energy is not entirely negligible.
This is shown in Fig.~\ref{fig:mfnew}, whose upper panel
compares the height of the barriers for the MF theory, the C2C model and the MF theory
using the C2C cutoff.
Up to $a=1$ the three curves are practically indistinguishable, but 
this is no longer true in the lower panel, where the corresponding parametric curves
are plotted.
The curves obtained by using the MF theory with the C2C cutoff have a double purpose:
to stress the secondary importance of the cutoff and to
justify the approximation made in Fig.~\ref{fig:mf6}, where we used the C2C cutoff to plot the MF results.
This trick greatly simplify the analysis in the MF case.

    The re-entrant character of curves in Figs.~\ref{fig:mf6} and \ref{fig:mfnew}
is a consequence of the modest bounding capacity of the potential, so that
    the grandcanonical theory is unable
    to give a correct description of the system. In some sense, it is a ``static'' proxy
    of the loss of metastability.

\section{Conclusions}
\label{sec.conclusions}

In this paper, we have developed a mean-field approach and thereby derived
a grandcanonical description of the DNLS model. This includes
the equilibrium state at any positive temperature and the homogeneous, metastable state
emerging for negative large temperatures.
The approximation consists in assuming that the interaction term is gauged by the average
mass (amplitude) of the neighbouring sites rather than by their actual value.
This approximation has no effect on either the critical curve (because the interaction term
is negligible) or the ground state curve (because all masses are equal). 
On the other hand, it is crucial for the phases to be handled accurately, as this approximation ensures.
In a certain sense, the mean-field approximation is a step forward compared to the 
stochastic C2C model in which the coupling term is completely neglected
and phase variables do not appear.

Our results show that the MF approximation provides a good-to-excellent description of the
system for all values of mass and energy densities via a direct grandcanonical description.
In particular, it provides an accurate characterization across the transition 
from positive to negative temperature, where instabilities are expected
to arise. In fact, while MF theories typically fail in proximity of phase transitions,
here the approximation works because the interaction energy is therein negligible,
and the transition itself occurs even in the absence of interactions.
Our approach shows that the presence of interactions keeps smooth the crossing of the critical curve.
The negative-$T$ equilibrium state cannot be described by the grandcanonical theory,
but for small, negative $\beta$, the system is trapped for a long time in a metastable state.
This passage is smooth and Eqs.~(\ref{eq.qM}-\ref{eq.hnlwM2}) are equally
valid for positive and negative temperatures.

The ensuing transition from the homogeneous metastable state to the localized 
equilibrium one goes beyond the scope of this article. It would require the
formulation of an appropriate dynamical rule consistent with the
mean-field approximation introduced in this paper.

	\section*{Acknowledgements}
	SI and PP acknowledge financial support from the Italian MUR PRIN2022 project ``Breakdown of
	ergodicity in classical and quantum many-body systems" (BECQuMB) Grant No.
	20222BHC9Z.

	\appendix

\section{Technical details for numerical simulations}
\label{app.tech}

Simulations of the full DNLS model are performed on a chain of size $N = 100$ with periodic boundary conditions
interacting with a thermal reservoir on site $n=1$.
The evolution is studied introducing the conjugate canonical variables 
\begin{equation}
    v_n = \sqrt{2 c_n}\cos{\phi_n}, \; u_n = \sqrt{2 c_n} \sin \phi_n, 
    \quad n = 1, \dots, N,
\end{equation}
whose time dependence is ruled by the following equations of motion
\begin{equation}
    \dot{v}_n = -(v_n^2 + u_n^2)u_n - u_{n-1} -u_{n+1}
\end{equation}
\begin{equation}
    \dot{u}_n = (v_n^2 + u_n^2)v_n + v_{n-1} +v_{n+1}.
\end{equation}
We have implemented this Hamiltonian dynamics by using a 4-th order Runge Kutta algorithm with an integration step $\delta t = 10^{-3}$.

The DNLS chain is thermostatted at temperature $T = \beta^{-1}$ and chemical potential $\mu$ with a Monte Carlo heat bath~\cite{iubini12}. 
Once every $t = 1$ units of time, we generate new canonical variables for site 1, $u_1' = u_1 + \delta u_1'$ and $ v_1' = v_1 + \delta v_1$,
where $\delta u_1$ and $\delta v_1$ are random variables extracted from the uniform distribution 
in the interval $[-0.5, 0.5]$.
Then, we update the canonical variables using a Metropolis algorithm by evaluating the cost function $W = \exp{(-\beta\Delta E + m \Delta A )}$, where 

\begin{equation}
    \Delta A = \frac{u_1'^2 + v_1'^2}{2} - \frac{u_1^2 + v_1^2}{2} = c_1' - c_1 
\end{equation}

\begin{equation}
    \Delta E = c_1'^2 - c_1^2 + (u_N +u_2)\delta u_1 + (v_N +v_2)\delta v_1
\end{equation}
are the variations in mass and energy respectively. When $W>1$, we update $u_1$ and $v_1$  with $u'_1$ and $v'_1$,
while if $W<1$ the move is accepted with probability $W$. This algorithm is implemented in the same way regardless of the sign of $\beta$~\cite{Baldovin_2021_review}.

Initial condition are typically selected as ``infinite temperature'' configurations, i.e. with phases $\phi_n$ extracted 
from the uniform distribution in $[0, 2\pi]$ and masses $c_n$ generated from an exponential distribution $f(c) = 1/a_i  \exp(-c/a_i)$,
with the $a_i$ chosen appropriately for each simulation so as to be close to the mass density reached at equilibrium.

The values of $a, h$ and $h_{nl}$ are obtained from a combination of spatial and temporal averages of local quantities.
With the canonical variables $u_n$ and $v_n$, the local masses are $c_n = \frac{1}{2} (u_n^2 + v_n^2)$. We can also define local energies $\e_n$ as
\begin{equation}
    \e_n \equiv \frac{1}{4} (u_n^2 + v_n^2)^2 + \frac{1}{2}(u_{n-1}+u_{n+1})u_n +\frac{1}{2}(v_{n-1}+v_{n+1})v_n,
\end{equation}
which correspond to the sum of the nonlinear local energy $c_n^2$ and half the coupling energy between sites $n-1$ and $n$ and between $n$ and $n+1$,
so that $\sum_{n=1}^N \epsilon_n = H$. 
We first consider for each site a time average of $c_n$, $\epsilon_n$ and $c_n^2$,
with which we make spatial averages to get $a$, $h$ and $h_{nl}$ respectively. Time averages were made over times between $10^7$ and $10^8$ time units,
after waiting for a transient $t = 10^5$. Only the 80 internal sites of the chain contribute to the spatial averages
(to remove any effects due to the heat bath, connected to site 1, the first and last 10 sites in the chain were excluded). 
The uncertainties of the averages are obtained from the standard deviation of the spatial average
and in all figures such uncertainties are either shown or they smaller than the symbol size.

\section{Small-$\beta$ expansion}
\label{app.exp}
In this appendix we derive the analytical expressions of the MF theory for small $\beta$.

Let us start with the case $\beta >0 $. The reduced grand canonical partition function of the MF model,
Eq.~(\ref{eq_z1}) $(J=1)$, 
\begin{equation}
    z = 2\pi \int_0^\infty d c  \, {\rm e}^{-\beta c^2 + mc} I_0(2 \beta q \sqrt{c})
    \label{eq:app_z1}
\end{equation}

can be rewritten as 
\begin{equation}
  z =  \int_{0}^\infty d c  \,\sum_{n = 0}^\infty {\rm e}^{(-\beta  c^2 +m c )}   \frac{2\pi}{(n!)^2}   \left(\beta q\right)^{2n} c^n,
  \label{eq:app_z2}
\end{equation}
by using the Taylor expansion of the modified Bessel function $I_0$,
\begin{equation}
    I_0 (v) = \sum_{n = 0}^\infty \frac{1}{(n!)^2} \left(\frac{v}{2}\right)^{2n}.
    \label{eq:app_I0}
\end{equation}
Since the series in the integrand of (\ref{eq:app_z2}) absolutely converges for any positive $c$, we can swap the order of integration and summation,
then individually integrate the terms of the series, from which we obtain

\begin{equation}
  z = \sum_{n = 0}^\infty 2^{-n}  \beta^{-(1+3n)/2}q^{2n} \pi\, U\left(\frac{1+n}{2}, \frac{1}{2}, \frac{m^2}{4\beta}   \right),
  \label{eq:app_z3}
\end{equation}
where $U$ is the Tricomi confluent hypergeometric function.

Similarly, it is possible to obtain the average value of any power $\alpha$ of $c$ as a function of $\beta$, $m$ and $q$,
\begin{equation}
\begin{split}
    \langle c^\alpha \rangle &=\frac{1}{z} \int_0^\infty d c \int_0^{2\pi} d \varphi \, c^\alpha \, {\rm e}^{-\beta c^2 + mc -2\beta q\sqrt{c } \cos \varphi }\\
    &= \frac{2\pi}{z} \int_0^\infty d c \, c^\alpha \, {\rm e}^{-\beta c^2 + mc} I_0(2 \beta q \sqrt{c})\\
    &=\frac{1}{z} \sum_{n = 0}^\infty2^{-n - \alpha} \pi \beta^{(n- 1 - \alpha)} (\beta q)^{2n} 
	\Gamma(n + 1 + \alpha) \\
	&\phantom{x} \times  U\left(\frac{n + 1 + \alpha}{2}, \frac{1}{2}, \frac{m^2}{4 \beta}\right),
\end{split}
\label{eq:app_caplha}
\end{equation}
from which we have the expression for the mass density $a$ for $\alpha = 1$ and for the nonlinear energy $h_{nl}$ for $\alpha = 2$. 
We can also express the self-consistency relation $q = \langle \sqrt{c}\rangle$ using Eq.(\ref{eq:app_caplha}) with $\alpha = 1/2$,

\begin{equation}
  \begin{split}
    q = \langle \sqrt{c} \rangle &=  \frac{\pi}{z} \sum_{n = 0}^\infty2^{-n - 1/2} \beta^{3n/2-3/4} \frac{q^{2n}}{(n!)^2} \Gamma\left(\frac{3}{2} + n\right)\\
     &\phantom{x} \times U\left(\frac{3}{4} + \frac{n}{2}, \frac{1}{2}, \frac{m^2}{4\beta}\right).
     \end{split}
\label{eq:app_scr}
\end{equation}

For the interaction energy density $h_{int}$ we have instead
\begin{equation}
\begin{split}
    h_{int} &=\frac{1}{z} \int_0^\infty d c\int_0^{2\pi} d \varphi \, \, 2 q \sqrt{c} \cos \varphi  \, {\rm e}^{-\beta c^2 + mc -2\beta q\sqrt{c } \cos \varphi }\\
    &= -\frac{4\pi q}{z} \int_0^\infty d c \, \sqrt{c} \, {\rm e}^{-\beta c^2 + mc} I_1(2 \beta q \sqrt{c}) \\
    &= \frac{1}{z}\sum_{n = 1}^\infty \frac{2^{-n} \beta^{-2 + 3n/2} m q^{2n}}{(n-1)!} U\left(1 + \frac{n}{2}  , \frac{3}{2}, \frac{m^2}{4 \beta} \right),
\end{split}
\label{eq:app_hint1}
\end{equation}
where we have used the expansion of $I_1$,
\begin{equation}
    I_1 (v) =  \frac{d}{d v} I_0 (v) =  \sum_{n = 1}^\infty \frac{n}{(n!)^2} \left(\frac{v}{2}\right)^{2n-1}.
    \label{eq:app_I1}
\end{equation}

It is now convenient to introduce the smallness parameter  $w = \beta/m^2$, which appears as the argument of the hypergeometric function $U$.  
In particular, $U(\cdot, \cdot, 1/4w)$ admits an asymptotic expansion for $w\rightarrow 0$ in terms of the generalized hypergeometric series $_2F_0$ \cite{Andrews_Askey_Roy_1999}, 
\be
U\left(a, b, \frac{1}{4w}\right) \simeq \left(4w\right)^{a} \,_2F_0 (a, a-b+1;;-4w),
\ee
where $_2F_0(a, a-b+1;;-4w)$ does not converge in $w = 0$ but exists as a formal power series in $w$.

From now on we will also consider $m<0$. By replacing $\beta = w m^2$ and expanding $U$ in powers of $w$ in Eq. (\ref{eq:app_z3}), we get
\begin{equation}
    z = \frac{2\pi}{|m|} - \frac{4\pi w}{|m|} + \left(\frac{24 \pi}{|m|} + 2 m^2 \pi q^2\right)w^2 + O(w^3)
    \label{eq:app_z4},
\end{equation}
that we can substitute in Eq. (\ref{eq:app_scr}) to obtain
\begin{equation}
    q = \frac{\sqrt{\pi}}{2 \sqrt{|m|}} - \frac{7\sqrt{\pi}}{8\sqrt{|m|} }w + \frac{\sqrt{\pi} (449 + 16 |m|^3 q^2) }{64 \sqrt{|m|}} w^2 + O(w^3, q^4).
\end{equation}

The self-consistency relation can therefore be solved in a perturbative way at an arbitrary order of $w$, from which we obtain

\begin{equation}
    q = \frac{\sqrt{\pi}}{2 \sqrt{|m|}} - \frac{7\sqrt{\pi}}{8\sqrt{|m|} }w + \frac{449 \sqrt{\pi}   + 16 |m|^2 \pi^{3/2} }{64 \sqrt{|m|}} w^2 + O(w^3).
    \label{eq.qq}
\end{equation}
Once we have $q$, we can replace it into the expressions $(\ref{eq:app_z4})$, $(\ref{eq:app_caplha})$ and $(\ref{eq:app_hint1})$ to get
\be
z = \frac{2 \pi}{|m|} - \frac{4 \pi }{|m|}w + \frac{\pi (24 + 2|m|^3 q^2)}{|m|} w^2 + O(w^3) 
\ee

\be
a = \frac{1}{|m|}-\frac{4 w}{|m|}+\left(\frac{40}{|m|}+\frac{|m| \pi }{4}\right) w^2 + O\left( w^3\right) 
\label{eq.aw}
\ee

\be
    h_{nl} = \frac{2}{m^2}-\frac{20 w}{m^2}+\frac{\left(296+m^2 \pi \right) w^2}{m^2} + O\left( w^3\right) 
\label{eq.hnlw}
\ee
\be
   h_{int} =  -\frac{\pi  w}{2}+\frac{15 \pi  w^2}{4} + O\left( w^3\right) 
\ee
\be
    h =  \frac{2}{m^2} -\left(\frac{\pi }{2} +\frac{20 }{m^2} \right) w+\left(\frac{19\pi }{4} +\frac{296 }{m^2} \right) w^2 + O\left( w^3\right) 
\label{eq.hw}
\ee

The combination of Eqs.~(\ref{eq.aw}) and (\ref{eq.hw}) gives, for $T=\infty$ (i.e., for $w=0$),
\be
h = 2a^2 .
\label{eq:hmfb0}
\ee

Let us now consider the case where $\beta<0$ and $m<0$. We can write the reduced partition function $z$ as
\be
z = 2\pi \int_0^{\frac{m}{2 \beta}} dc \, {\rm e}^{-\beta c^2 + m c} I_0(2 \beta q \sqrt{c}),
\ee
where the cutoff in the mass integration $c^\ast =\frac{m}{2\beta}$ is the same as that assumed in the study of
the C2C model.

If we express $\beta$ as $\beta = w m^2$ and expand the integrand function in powers of $w$ up to order $w^2$, we obtain

\begin{equation}
\begin{split}
        z &\simeq 2\pi \int_0^{\frac{m}{2\beta}} d c\, {\rm e}^{m c} \left( 1 - c^2 m^2 w + c m^4\left(\frac{c^3}{2} + q^2\right) w^2\right)\\ 
        z&= \frac{2 \pi}{|m|} - \frac{4 \pi }{|m|}w + \frac{\pi (24 + 2|m|^3 q^2)}{|m|} w^2 + O(w^3),
\end{split}
\end{equation}
which is equal to Eq. (\ref{eq:app_z4}). We therefore find that the expansion of $z$ at negative temperatures coincides with that at $\beta >0$. 
An adjustment of the cutoff leads only to corrections of order ${\rm e}^{-1/(2|w|)}$.

Similarly, from the calculation of $q$, $a$, $h_{nl}$ and $h_{int}$ we find that the expressions (\ref{eq.qq}-\ref{eq.hw}) are also valid for $\beta<0$.

\begin{figure}
        \centering
        \includegraphics[width=1\columnwidth]{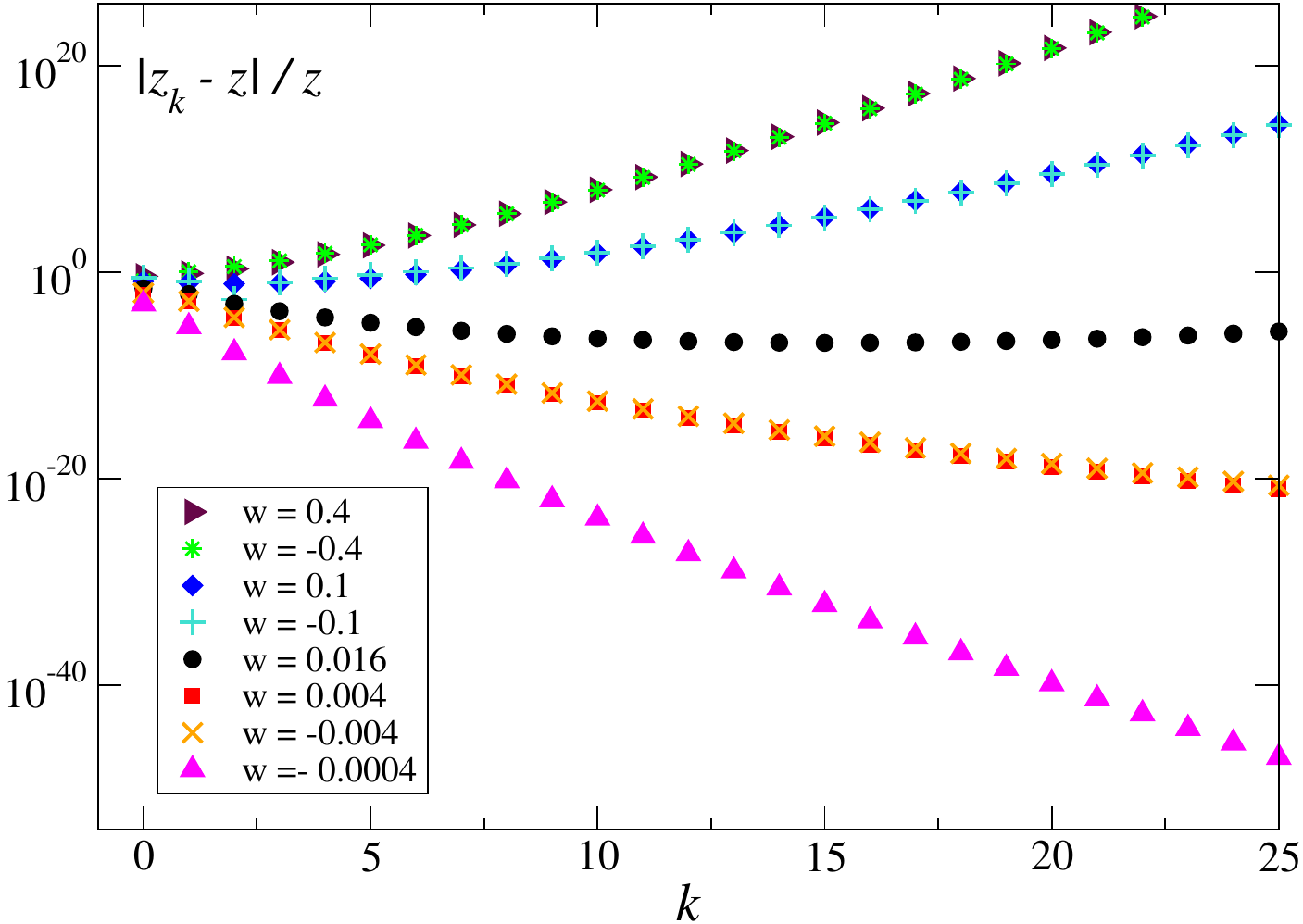} 
	    \caption{
		    Relative difference between the MF partition function $z$ computed by numerically integrating the exact expression
                    (\ref{eq_z1}) and its analytical approximation $z_k$ obtained as a series expansion at order $k$ 
                    in $w$. The various symbols correspond to different $w$ values as from the legend. The value of $\beta$ is
                    always equal to $\pm 0.4$ except for the lowermost curve, where $\beta=-0.01$. Notice that the sign of $\beta$ 
                    coincides with the sign of $w$.
	    For negative $\beta$, the numerical integration is performed by imposing the same cutoff $c^*$ as in the C2C model.
        }
        \label{fig:mf10}
    \end{figure}

Higher-order terms can be obtained and determined with the help of Mathematica software up to the
desired accuracy.
In Fig.~\ref{fig:mf10} we report the relative error on the MF-value 
of the single-particle partition function 
$z(\beta,\mu)$ versus the approximation order $k$, for some values of $w$ (both for positive and negative
temperatures).

As expected, the very high accuracy for small $w$ values degrades when $w$ becomes of order 1.
A second feature is that the accuracy of the expansion initially increases with $k$ up to some $k^*$,
when it starts degrading.  This is a clear indication of the asymptotic nature of the series expansion in powers of $w$.
Finally, the overlap of data obtained for the same $w$ but different sign of $\beta$ 
indicate that the sign is irrelevant.

\section{$T=0$ limit}
\label{app.T0}

In the low-temperature limit, we can approximate the expressions for the partition function $z$, $a$, $q$ and $h$ by using the leading term of the modified Bessel functions $I_{0,1}$ for large argument $v$
\be
    I_{0,1}(v) \simeq \frac{{\rm e}^{v}}{\sqrt{2 \pi v}} .
    \label{eq:App_I01}
\ee
By substituting $(\ref{eq:App_I01})$ in Eq. $(\ref{eq_z1})$, $z$ can be rewritten as 
\be
    z \simeq \int_0^\infty dc \sqrt{\pi} \frac{{\rm e}^{\beta A(c) } }{\sqrt{\beta q } c^{1/4}},
\ee
where
\be
    A(c) = -c^2 + \mu c + 2 q \sqrt{c}.
\ee
We can now use the saddle-point method to approximate $z$ as  
\be
    z \simeq \sqrt{\pi} \frac{{\rm e}^{\beta A(c^\ast) } }{\sqrt{\beta q } c^{\ast1/4}},
\ee
where $c^\ast$ is the mass value that maximizes $A(c)$ and therefore satisfies the relation
\be
\frac{\partial A(c)}{\partial c}\bigg|_{c=c^\ast} = 2 c^\ast - \frac{q}{\sqrt{c^\ast}} - \mu =    0.
\label{eq:App_c*}
\ee

We can also derive the expression for $q$ in the low-temperature limit,
\be
      q \simeq \frac{1}{z} \int_0^\infty d c\, \sqrt{c} \sqrt{\pi} \frac{{\rm e}^{\beta A(c) } }{\sqrt{\beta q} c^{1/4}} \simeq \sqrt{c^\ast} \frac{\sqrt{\pi}}{z}   \frac{{\rm e}^{\beta A(c^\ast) } }{\sqrt{\beta q } c^{\ast1/4}} = \sqrt{c^\ast},
      \label{eq:App_q}
\ee
and similarly for the expressions for $a$ and $h$
\be
      a \simeq \frac{1}{z} \int_0^\infty d c\, c \sqrt{\pi} \frac{{\rm e}^{\beta A(c) } }{\sqrt{\beta q } c^{1/4}} \simeq c^\ast,
      \label{eq:App_a}
\ee
\be
  \begin{split}
      h &\simeq \frac{1}{z} \int_0^\infty d c\, (c^2 - 2 q \sqrt{c}) \sqrt{\pi} \frac{{\rm e}^{\beta A(c) } }{\sqrt{\beta q} c^{1/4}}\\
      &\simeq c^{\ast2} - 2\sqrt{c^\ast }q  \simeq a^2 - 2a.  
      \label{eq:hmfT0}
  \end{split}
\ee
Analogously, we find $h_{nl} \simeq a^2$ and $h_{int}\simeq -2a$.

If now we substitute (\ref{eq:App_q}) and (\ref{eq:App_a}) in Eq. (\ref{eq:App_c*}), we obtain 
\be
    c^\ast = a = \frac{\mu + 1}{2},
    \label{eq:App_amuMF}
\ee
that is, a relation between $a$ and the chemical potential $\mu$ for the MF model.

For the complete DNLS model, using the saddle-point method we obtain a different relation between $a$ and $\mu$ in the low-temperature limit

\be
    \mu^{(T =0)}_{DNLS} = 2(a - 1)
\ee
and if we replace $a$ with Eq. $(\ref{eq:App_amuMF})$ we obtain a relationship between the chemical potentials in the two models 
\be
    \mu_{MF}^{(T =0)} =  \mu^{(T =0)}_{DNLS} +1.
\ee

\section{The ratio of energy differences $R$}
\label{app.Hlnl}

If we combine Eqs.~(\ref{eq.aw}-\ref{eq.hw}) retaining all terms up to the order $w$, we find
\bea
\Delta h_{nl} &\equiv & h_c - h_{nl} = \frac{4w}{m^2} \\
\Delta h &\equiv & h_c - h = \frac{4w}{m^2} + \frac{\pi w}{2} ,
\eea
so that
\be
R \equiv \frac{\Delta h_{nl}}{\Delta h} = \frac{4}{4 + \pi m^2 /2} 
\ee
which tends to $8a^2/\pi$ for small $a$ and to 1 for large $a$.

\section*{References} 
\bibliographystyle{iopart-num}

\providecommand{\newblock}{}

\end{document}